\documentstyle[aps,preprint]{revtex}
\def\gsim
 {\raise2pt\hbox{$\displaystyle{}\mathrel{\mathop>_{\raise1pt\hbox{$\sim$}}}
{}$}}
\def\lsim
 {\raise2pt\hbox{$\displaystyle{}\mathrel{\mathop<_{\raise1pt\hbox{$\sim$}}}
{}$}}
\begin{document}
\title{Macroscopic Quantum Tunneling of a Domain Wall in a Ferromagnetic Metal}
\author{Gen Tatara and Hidetoshi Fukuyama }
\address{        Department of Physics, University of Tokyo,
        7-3-1 Hongo, Tokyo 113, Japan }
\date{\today}
\maketitle
\newcommand{\beq}{\begin{equation}}
\newcommand{\eeq}{\end{equation}}
\newcommand{\beqa}{\begin{eqnarray}}
\newcommand{\eeqa}{\end{eqnarray}}
\newcommand{\sigbf}{\mbox{\boldmath$\sigma$}}
\newcommand{\Mv}{{\bf M}}
\newcommand{\nv}{{\bf n}}
\newcommand{\Jv}{{\bf J}}
\newcommand{\kv}{{\bf k}}
\newcommand{\qv}{{\bf q}}
\newcommand{\xv}{{\bf x}}
\newcommand{\half}{\frac{1}{2}}
\newcommand{\kf}{k_{\rm F}}
\newcommand{\kfu}{k_{{\rm F}\uparrow}}
\newcommand{\kfd}{k_{{\rm F}\downarrow}}
\newcommand{\kfs}{k_{{\rm F}\sigma}}
\newcommand{\ef}{\epsilon_{\rm F}}
\newcommand{\ek}{\epsilon_{\bf k}}
\newcommand{\efs}{\epsilon_{{\rm F}\sigma}}
\newcommand{\eks}{\epsilon_{{\bf k}\sigma}}
\newcommand{\ekqs}{\epsilon_{\kv+\qv,\sigma}}
\newcommand{\ekd}{\epsilon_{\kv\downarrow}}
\newcommand{\ekqu}{\epsilon_{\kv+\qv,\uparrow}}
\newcommand{\fks}{f_{\kv\sigma}}
\newcommand{\fkqs}{f_{\kv+\qv,\sigma}}
\newcommand{\fkd}{f_{\kv\downarrow}}
\newcommand{\fkqu}{f_{\kv+\qv,\uparrow}}
\newcommand{\cel}{c_{\xv\sigma}}
\newcommand{\ael}{a_{\xv\sigma}}
\newcommand{\cd}{c_{\xv\sigma}^\dagger}
\newcommand{\ad}{a_{\xv\sigma}^\dagger}
\newcommand{\thx}{\theta_{\xv}}
\newcommand{\phx}{\phi_{\xv}}
\newcommand{\Hu}{H_{\rm U}}
\newcommand{\Hus}{H_{\rm U}^{(\rm slow)}}
\newcommand{\Huf}{H_{\rm U}^{(\rm fast)}}
\newcommand{\Ms}{M}
\newcommand{\Cpm}{C_{+-}}
\newcommand{\Csig}{C_{\sigma}}
\newcommand{\oml}{\omega_\ell}
\newcommand{\chiu}{\chi^0_{\uparrow}}
\newcommand{\chid}{\chi^0_{\downarrow}}
\newcommand{\DSone}{\Delta S^{(1)}}
\newcommand{\DStwo}{\Delta S^{(2)}}
\newcommand{\nzero}{n_0}
\newcommand{\Awall}{A_{\rm w}}
\newcommand{\Spin}{S}
\newcommand{\ktil}{\tilde k_0}
\newcommand{\lamtil}{\tilde \lambda}
\newcommand{\Vtil}{ V}
\begin{abstract}
The macroscopic quantum tunneling of a planar domain wall in a ferromagnetic
metal is studied based on the Hubbard model.
It is found that
the ohmic dissipation is present even at zero temperature due to the
gapless Stoner excitation, which is the
crucial difference from the case of the insulating magnet.
The dissipative effect is calculated as a function of width of the wall
and is shown to be effective in a thin wall and in a weak ferromagnet.
The results are discussed in the light of recent experiments on
ferromagnets with strong anisotropy.
\end{abstract}
%
%

In recent years, owing mainly to the development of technology in
mesoscopic physics,
there has been growing interest in macroscopic quantum tunneling (MQT)
 in magnetic systems\cite{SCB}, e.g., the magnetization reversal in small
grains\cite{CG},
the quantum nucleation of  a domain\cite{CGnuc}, and  the
quantum depinning of a domain
wall via MQT\cite{Sta}.
These studies are mainly in ferromagnets, but recently
a magnetization reversal due to MQT has been observed
in antiferromagnetic particles of horse spleen ferritin\cite{ASG}.
In the case of the quantum depinning of a
 domain wall pinned by defects,
the position of the wall at the pinning center becomes metastable in the
external magnetic field, and
if the barrier height is low enough, the position tunnels
out of the local minimum.
This problem was studied theoretically by Stamp\cite{Sta} for the case of
an insulating magnet.
The tunneling rate was expressed in terms of macroscopic variables,
and was shown to  be large enough to be observed
even for a large wall with about $10^{10}$ spins.
As sources of dissipation, which is shown to be important by the seminal
paper by Caldeira and Leggett\cite{CL},
 Stamp considered  magnons and phonons,
but the effects turn out to be negligible, since magnon has a gap
and coupling to phonon is
weak.
Consequently it has been concluded that the tunneling rate
is not essentially affected by dissipation in insulators.

Experiments on MQT in magnetic system, however, have been carried out in
metallic ferromagnets.
In metals, in contrast to the case of insulators, there is a gapless
excitation of spin flip,
and hence dissipation from conduction electrons must be very
important.
Consequently the quantum motion of the wall in metals should be quite different
from that in insulators.\cite{CIS}
In this paper, we will investigate theoretically the dissipative effect
on MQT of a domain wall in a ferromagnetic metal based on an itinerant electron
model.
An important and interesting feature of the itinerant system is that
the electron, which supports magnetization, works also as a source
of dissipation in the dynamical motion of the magnetization itself.
Our analysis  is based on the Hubbard model in the continuum.
The calculation is carried out at zero temperature,
since we are interested only in the quantum tunneling present at low
temperature.

The Lagrangian in the imagnary time path integral is given by
\begin{equation}
L=\sum_{\kv\sigma} c^\dagger_{\kv\sigma}
(\partial_\tau+\epsilon_\kv) c_{\kv\sigma}
   + U\sum_{\xv} n_{\xv\uparrow}n_{\xv\downarrow}  , 		\label{L0}
\end{equation}
where $\cel$ is an electron operator at site $\xv$ with spin $\sigma(=\pm)$,
 $n_{\xv\sigma}\equiv\cd\cel$ and $U$ is the Coulomb repulsion.
The band energy is $\epsilon_\kv\equiv \kv^2/(2m)-\ef$
with the fermi energy $\ef$.
The Coulomb repulsion term will be rewritten by introducing a
Hubbard-Stratonovich
field representing the magnetization; $\Mv_\xv\equiv M_\xv \nv_\xv$, where
$M_\xv\equiv
<(c^\dagger {\bf \sigbf} c)_\xv >  \nv_\xv $ with $\nv_\xv$ being a slowly
varying unit vector which describes the direction of magnetization.
The magnitude of magnetization is assumed as space-time
independent, $M_\xv\equiv\Ms$.
Hence
only $\nv_\xv$ remains as the relevant degree of freedom.

The spatial variation of $\nv_\xv$ accompanied with a domain wall is assumed
to be much slower compared to
the inverse fermi momentum of the electron $\kf^{-1}$.
For the analysis of such a slowly varying field,
a locally rotated frame\cite{Pra} of electron is convenient such that
the $z$-axis of the
electron is chosen in the direction of the local magnetization vector
$\nv_\xv$.
The electron operator $a_{\xv\sigma}$ in the new frame  is related to the
original $\cel$ as
\beq
 a_{\xv\sigma} \equiv \sigma \cos({\theta}/{2})c_{\xv\sigma} +
  e^{- i\sigma\phi}\sin({\theta}/{2})c_{\xv,-\sigma}
\label{loc}
\eeq
where the polar coordinates $(\thx(\tau),\phx(\tau))$ parametrize the
direction of $\nv_\xv(\tau)$.
The electron $a_{\xv\sigma} $ is polarized uniformly with the energy
$\epsilon_{{\kv}\sigma}\equiv {{\kv}^2}/{2m}-\sigma UM-\ef$.
As a price of this transformation, there arises from the kinetic term
$c^\dagger \dot c + |\nabla c|^2/(2m)$ an additional term $H_{\rm int}$
that describes the interaction of electrons
with spatial variation of the magnetization vector\cite{Pra}.
This interaction $H_{\rm int}$ is small and of the order of
$O(\kfu\lambda)^{-1}$, where $\lambda$ is the domain wall thickness,
$\kfu$ is the fermi  momentum of the majority spin,
 and hence can be treated perturbatively.
Our following results are valid for $\lambda\kfu\gsim 1$.

The integration over the electron degrees of freedom leads to the
effective action for the magnetization  as
$
S_{\rm eff}=
-{\rm tr}\ln(\partial_\tau+\epsilon_{\kv\sigma})+\beta\sum_\xv({U}/{2})M^2
+\Delta S.
$
The first two terms are the mean field action for a ferromagnet
which determines the magnetization $M$.
The dynamics of $(\theta,\phi)$ is described by
$\Delta S$, which is expressed in terms of correlation functions of electron.
This term is decomposed into two parts, that is local and non-local in time,
respectively,
as $\Delta S\equiv \Delta S_{\rm loc}+\Delta S_{\rm dis}$.
The local part $ \Delta S_{\rm loc}$ determines the dynamics of
magnetization vector, and the non-local part $\Delta S_{\rm dis}$ represents
the dissipative effect due to conduction electrons on the motion of
the magnetization vector.

Up to the lowest order in $\partial_\tau$ and $\nabla$,
the local part $\Delta S_{\rm loc}$ turns out to be formally
 the same as the ferromagnetic Heisenberg model\cite{Pra,RS}
with spin $S\equiv \Ms/2$ whose Lagrangian is given by
\beq
L_{\rm H}= \int {d^3\xv} \left[ i\frac{S}{a^3}\dot{\phi}(1-\cos\theta)+
   \frac{JS^2}{2}\left( (\nabla\theta)^2+\sin^2\theta(\nabla\phi)^2 \right)
\right]. \label{Sloc}
\eeq
The exchange coupling or the spin wave stiffness constant is expressed by
the parameters of the original Hubbard model as
$
J\equiv({n}/{ma^3\Ms^2})\left[1-({(\kfu^{5}-\kfd^5)a^3})/({30\pi^2 mnU\Ms})
\right] $,
where $n$ is the electron number per site,
$\kfs\equiv(2m(\ef+\sigma UM))^{1/2}$
is the fermi momentum and $a$ is the lattice constant.
Hence, in the absence of the non-local term $\Delta S_{\rm dis}$,
there is no formal difference between metallic and insulating ferromagnets,
and the tunneling rate of the domain wall  is determined on the same
footing\cite{Sta}.

In order to incorperate the domain wall, the anisotropy energy
 with $yz$  easy plane is introduced\cite{aniso};
\beq
H_{\rm ani}= \int d^3\xv\left(-\frac{K}{2}S^2\cos^2\theta
+\frac{K_\perp}{2}S^2\sin^2 \theta\cos^2\phi \right).
\eeq
The Lagrangian $L_{\rm H}+H_{\rm ani}$  has a planar domain wall centered
at $x=Q(\tau)$ and moving slowly as a classical solution;
$\cos\theta(\xv, \tau) = \tanh ({x-Q(\tau)}/{\lambda})$ and
$\cos\phi(\xv,\tau) \simeq i{\dot{Q}}/{c} \ll 1$
with $c\equiv K_\perp \lambda S a^3$ where
$\lambda\equiv \sqrt{{J}/{K}}$ is the width of the wall.
This configuration is depicted in Fig. \ref{DW}.

For the magnetic field $H$ close to the coercive field $H_c$, {\it i.e.},
$(H_c-H)/H_c\equiv\epsilon\ll1$,
the potential for the wall coordinate $Q$ is given by
$V(Q)\equiv (1/2)M_w \omega_0^2 Q^2 [1-(Q/Q_0)^2]$
where $M_w\equiv 2N/(K_\perp \lambda^2 a^3)$ is the domain wall mass
$N$ being the number of the spins in the wall.
For this case of small $\epsilon$, the
attempt frequency around the minimum is
$ \omega_0 \simeq (\mu_0(\hbar\gamma)^2/a^3)  \sqrt{h_c}\epsilon^{\frac{1}{4}}$
and the width of the barrier is given by
$Q_0 = \sqrt{{3}/{2}}\sqrt{\epsilon}\lambda$ where
 $h_c\equiv H_c/(\hbar\gamma S /a^3)$ is the ratio of the coercive field
to the magnetic moment per unit volume
 ($\mu_0$ is the magnetic peameability of free space  and  $\gamma$ is
the gyromagnetic
ratio).
The actual value of attempt frequency is
$\omega_0 \simeq 5\times \sqrt{h_c}\epsilon^{\frac{1}{4}} \,\, ({\rm K})$
for the choice of $a=3$\AA,
and in the present case, this is roughly the same as
the crossover temperature $T_{co}$ from the thermal activation
to the quantum tunneling.
The classical solution (bounce) of $Q$ in the metastable potential $V(Q)$
is given by
$Q(\tau)=Q_0/{\cosh^2({\omega_0\tau}/{2})}$, and
the tunneling rate out of the local minimum is estimated by use of this bounce
solution.
For the case of the wall with the cross sectional area $Na^3/\lambda$ as
shown in Fig. \ref{DW},
the rate $\Gamma_0$ without dissipation is reduced to
$\Gamma_0=A\exp({-B})$ where
$A\simeq (\mu_0(\hbar\gamma)^2/a^3)N^{1/2}h_c^{{3}/{4}}\epsilon^{{7}/{8}}
\simeq  10^{11} N^{1/2}h_c^{{3}/{4}}\epsilon^{{7}/{8}} \,\,({\rm Hz})$
and the exponent,$B$, is given
as $B\simeq N h_c^{1/2}  \epsilon^{{5}/{4}}$.\cite{Sta}
Since  $B$ is proportional to $N\epsilon^{{5}/{4}}$\cite{exponent},
a  small value of $\epsilon$
is needed to observe the tunneling.

Let us now look into the non-local action $\Delta S_{\rm dis}$,
where the characteristic feature of the itinerant electron system is to
be seen.
For the case of a weak dissipation, this contribution is evaluated
by use of the configuration of a domain wall obtained in the absence
of dissipation.
Up to $\nabla^2$,
$\Delta S_{\rm dis}$ is obtained as
\beqa
\lefteqn{  \Delta S_{\rm dis}=
  \frac{1}{(4m)^2}\int\! d\tau \! \int \! d\tau'\frac{1}{\beta}
\sum_{\ell} e^{i\oml(\tau-\tau')}\sum_{i} \sum_\qv q_i^2
 } \nonumber\\
&&
 \times |\theta_\qv(\tau)-\theta_\qv(\tau')|^2
 <J_+^i(\qv) J_-^i(-\qv)>|_{i\oml},
 \label{DS}
\eeqa
where $\Jv_\alpha (\qv)$ $(\alpha=\pm,z)$ are the Fourier transform of
the spin currents of the electron;
$
\Jv_\alpha \equiv -i[(a^\dagger\sigma_\alpha \nabla a) -
(\nabla a^\dagger\sigma_\alpha a)] $
with $\sigma_\pm\equiv\sigma_x\pm i\sigma_y$, and
$\theta_\qv\equiv \sum_\xv e^{-i\qv\xv}\theta_\xv$.
The dissipation does not result from  the $z$-component $J_z$ in the present
case of a domain wall motion with $\nabla \phi=0$.
The expectation value of  electron  spin
current $<J_+J_{-}>$ in $\Delta S_{\rm dis}$
is evaluated
by the random phase approximation (RPA)
 in the background of uniform magnetization.\cite{RPA}

After the analytic continuation to real frequency,
 $\Delta S_{\rm dis}$ is expressed by
the imaginary part of the retarded correlation function
$<J_+ J_->|_{\omega+i0}$ as\cite{ATF}
\beqa
\lefteqn{  \Delta S_{\rm dis}=
  \frac{1}{(4m)^2}\int\! d\tau \! \int \! d\tau'
 \sum_\qv q_x^2 |\theta_\qv(\tau)-\theta_\qv(\tau')|^2
 } \nonumber\\
&&
 \times \int_0^\infty \frac{d\omega}{\pi} e^{-\omega|\tau-\tau'|}
{\rm Im} <J_+^x(\qv) J_-^x(-\qv)>|_{\omega+i0}.
\eeqa
The imaginary part is expanded in terms of $\omega/\ef$ as
\beqa
{\rm Im}<\Jv^x_+(\qv) \Jv^x_-(-\qv)>|_{\omega+i0}=
  \left\{ \begin{array}{ccc}
   \omega\frac{m^2a^3(\kfu^2-\kfd^2)^2}{\pi|q|^3}  +O(\omega^3)
   &  &   \kfu-\kfd<|q|<\kfu+\kfd  \\
  0 &  & {\rm otherwise}.  \end{array}\right.
\label{Im}
\eeqa
The term linear in $\omega$ gives rise to the ohmic dissipation.
It is seen from the restriction on $q$ that the ohmic dissipation
arises from the Stoner excitation,
which is a gapless excitation of spin flip across the fermi energy.

By the expression of the domain wall configuration, the
non-local part of the effective action  is reduced to
\beqa
\lefteqn{ \Delta {S}_{\rm dis}=
N\frac{(\kfu^2-\kfd^2)^2a^4}{4}\frac{1}{\lambda a}
\int d\tau\int d\tau' \frac{1}{(\tau-\tau')^2}
 }\nonumber\\
&&\times\int^{\kfu+\kfd}_{\kfu-\kfd}\frac{dq}{2\pi}
\sin^2\frac{q}{2}(Q(\tau)-Q(\tau'))
\frac{ 1}{q^3} \frac{1}{\cosh^2 \frac{\pi}{2}\lambda q} .\label{DSdis}
\eeqa
The form factor of the wall, $1/\cosh ^2 (\pi\lambda q/2)$, represents the
effective coupling between electrons and the wall, and because of this factor,
the momentum integration is dominated by
$q\lsim\lambda^{-1}$.
The time integral is estimated by approximating the bounce solution
as $Q(\tau)\simeq Q_0 \theta(\omega_0^{-1}-|\tau|)$
and by introducing a
short time cutoff of $\omega_0^{-1}$ for the relative time
$(\tau-\tau')$\cite{KP}.
Noting  $q Q_0\propto q \lambda \sqrt\epsilon \ll 1$,
the sine function in Eq. (\ref{DSdis}) can be replaced by its argument and
 the action is evaluated to be
$\Delta  S_{\rm dis}\equiv \eta N\epsilon $
where the factor, $\epsilon$, is due to the smallness of the squared
tunnel distance $Q_0^2$.
Here the strength of dissipation, $\eta$, is
\beq
\eta= \frac{3\ln 3}{16\pi}{(\kfu^2-\kfd^2)^2 a^4}\frac{\lambda}{ a}
\int^{(\kfu+\kfd)\frac{\pi}{2}\lambda}_{(\kfu-\kfd)\frac{\pi}{2}\lambda}
dx\frac{1}{x}
\frac{1}{\cosh^2 x}  .\label{eta}
\eeq
For a thick wall $\lambda(\kfu-\kfd)\gg1$,
$\eta\propto \exp[-\pi\lambda(\kfu-\kfd)]$ and then the dissipation
is negligible.
On the other hand, $\eta$ can be large if
 $(\kfu-\kfd)\lambda \lsim 1$.
This condition is compatible with that of
slow spatial variation
$\lambda\kfu\gsim1$ for a wall with moderate thickness in a weak ferromagnet
and for a thin wall in a stronger ferromagnet.
The strength $\eta$ is plotted as a function of $(\lambda/a)$ in
Fig. \ref{FIGeta} for three different values of
$\delta\equiv(\kfu-\kfd)/(\kfu+\kfd)$
with $(\kfu+\kfd)a=6$ which may represent the case of an iron.
The dissipation is larger for weaker magnet (smaller $\delta$).
(For a complete ferromagnet, $\kfd$ vanishes and
the ohmic dissipation disappears.)
It is seen that $\eta$ can be of the order 0.1 for a wall with thickness a few
times the lattice spacing with $\delta\lsim 0.1$.
In the presence of dissipation, the tunneling rate is reduced to be
$\Gamma=A\exp[-(B+\Delta S_{\rm dis})]=\Gamma_0 \exp(-\eta N\epsilon)$.
Because of the different $\epsilon$-dependence of $B$ and
$\Delta S_{\rm dis}$, the ratio
$\Delta S_{\rm dis}/B=\eta h_c^{-1/2} \epsilon^{-1/4}$ is much larger
than unity for the case of small $\epsilon$ we are interested in,
and in particular for a thin wall
($h_c$ is usually small, e.g.,$\simeq 10^{-4}$).
Consequently the tunneling
rate is predominantly determined by dissipation in such cases.
The tunneling rate $\Gamma$ is
shown in  Fig. \ref{FIGgamma} for the case of insulator $(\eta=0)$ and
the typical case of a metal ($\eta=0.1$) by the broken and solid lines,
respectively for a choice of $h_c=10^{-4}$.
In this figure, the number of spins is taken either  $N=10^4$ or $10^8$.
The value $N=10^4$ corresponds, for instance, to a wall with thickness of
about $10$\AA\ and
the area of $200$\AA$\times 200$\AA.
The tunneling rate is seen to be much smaller in metals than that
in insulators.

We have neglected the effect of magnetic field on electronic states.
This is justified as long as $UM\gg\gamma H$.
In experimental situations with the magnetic field of $\lsim1$T and
$U\simeq 10$eV,
this condition reduces to $M\gsim 10^{-4}$ in unit of the Bohr magneton,
which is easy to satisfy.
However, in order to discuss the case of very small $M$, the fluctuation
of the magnitude $M_\xv$ around the mean field value must also be included.

The contributions of higher order in $H_{\rm int}$ are smaller than that of
the second order we have calculated; for the potential
renormalization by the order of $(\kfu\lambda)^{-2}$ and
for the dissipative effect
by $(\kfu\lambda)^{-2}$ or $\epsilon$.

In Eq. (\ref{eta}) we have taken account of only the ohmic dissipation.
The super-ohmic contributions, which are of higher orders of
$(\omega/\ef)$ in Eq. (\ref{Im}), are
smaller than the ohmic one by a factor of
$(\omega_0/\ef)^2 \ll 1$ and hence are negligible.
On the other hand, a contribution
from the magnon pole, which has not been taken into account
in the correlation function $<JJ>$,
 is calculated from
\beq
{\rm Im}<\Jv^x_+(\qv) \Jv^x_-(-\qv)>|_{\omega+i0}^{\rm (pole)}\simeq
\pi M^3 (Jmq)^2 \delta(\omega-\omega_q)
\eeq
where $\omega_q\equiv \Delta_0 +JMa^3q^2/2$ is the magnon energy with
the anisotropy gap $\Delta_0$.
This pole leads to super-ohmic dissipation, whose
 strength, $\eta^{\rm (pole)}$, is evaluated
as $\eta^{\rm (pole)}=(28/5)M(\Delta_0/\omega_0) \exp(-\Delta_0/\omega_0)$.
Since experiments are usually carried out in highly anisotropic materials with
 $\Delta_0/\omega_0\simeq 10$, this contribution is very small
compared to the ohmic dissipation for the case of a thin wall.

The present metallic case, where the ohmic dissipation is present even
at absolute zero, are in contrast with  the insulating case.
At finite temperatures, however,
there are ohmic dissipations even in the latter case.
Stamp calculated such ohmic dissipations from
two- and three-magnon processes and found
$\eta^{\rm (mag)}=$$(3/2\pi S)$$(1/\beta\Delta_0)$$\exp(-\beta \Delta_0)$.
These processes corresponds to  higher order contribution  of $H_{\rm int}$
in our calculation.
In contrast to the case of metals,
the value of $\eta^{\rm (mag)}$ vanishes at $T=0$ and is very small at
$\beta\Delta_0 \gg1$, hence the ratio of $\Delta S_{\rm dis}/B$ is negligibly
small in insulators.

In metals, eddy currents may influence MQT.
An electric field is induced by Faraday's law from the change of
magnetizations accompanied with the motion of the wall.
This field produces the electric current and thus leads to the Joule heat
of $P=(\mu_0 \hbar \gamma/a^3)^2\dot{Q}^2/\rho$ per unit volume
where $\rho$ is the resistivity.
By use of the specific heat $C$ and the system size of $L$, the temperature
rise due to the eddy current is expressed as $\Delta T=PL/(C \dot{Q})$.
For $\rho\simeq 10^{-7}$[$\Omega$m], $C\simeq10$[J/Km$^3$], and
$\dot Q\simeq1$[m/s] with $L=100$\AA, it is estimated as
$\Delta T\simeq 10$mK.
This value is not negligible but effects associated with this temperature
rise may be separated from the
intrinsic effects in careful experiments.

Our result shows a distinct difference between MQT of thin
walls in metallic and insulating magnets.
Unfortunately the experiments carried out so far appears not yet be able to
observe dissipation due to itinerant electrons.
In the experiment of a domain wall motion
in a small particle of Tb$_{0.5}$Ce$_{0.5}$Fe$_2$, MQT was observed below
$T_{co} \simeq 0.6$K\cite{PSB}.
However, the width of the domain wall is about $30$\AA\ and according to
our result, $\eta\propto \exp[-\pi\lambda(\kfu-\kfd)]$,
the dissipation from electron is negligible for such a thick wall.
This may be the reason why the result of the crossover temperature
$T_{co}\sim0.6$K is roughly in agreement  with the theory\cite{Sta}
  without dissipation.
On the other hand, the domain wall in SmCo$_5$ is
very thin $\lambda\simeq12$\AA, and our
result suggests strong effect of dissipation, which will be interesting
to observe.
Experiments on bulk crystal of
SmCo$_{3.5}$Cu$_{1.5}$ with very thin walls (a few times $a$) have
been performed\cite{UB},
although quantitative argument
is not easy since many walls participate in these experiments.
Even in the case of thick walls, the dissipative effect becomes large in
weak ferromagnets, where the experiments, however, will not be easy
because of small value of saturation magnetization $M$.

MQT in disordered magnets has a new  possiblity of observing a
significant effect of sub-ohmic dissipation.
In fact, as disorder is increased in a metallic magnet, the Anderson
transition
into an insulator will occur, and it was shown recently that near the
transition the  dissipation  due to the
conduction electron becomes sub-ohmic\cite{ATF}.
Disordered magnets may also be suitable for study of MQT because
the  eddy current becomes less important
for larger resistivity.

Our calculation is valid in the $s$-$d$ model as well, where
localized magnetic moment is due to $d$ electron and the current
is carried by $s$ electron.

In conclusion, we have studied the macroscopic quantum tunneling of a domain
wall in a metallic ferromagnet on the basis of Hubbard model.
The crucial difference from the case of an insulator is the presence of
ohmic dissipation even at zero temperature due to the
gapless Stoner excitation.
The coupling of domain wall to electrons is effective only for  momentum
transfer of $|q|\lsim\lambda^{-1}$, while
Stoner excitation is gapless at the restricted region
 $\kfu-\kfd<|q|(<\kfu+\kfd)$.
Hence
the effect is negligible for a thick domain wall
in which experiments so far have been
carried out.
On the other hand, important effects of the ohmic dissipation are expected in
thinner domain walls and in weak ferromagnets, which will be within the
present experimental attainability.

The authors are grateful to  H. Kohno, H. Yoshioka and M. Hayashi for
valuable discussions at every stage of this work.
G. T. also thanks K. Nosaka for her assistance in collecting articles.
This work is financially supported by Ministry-Industry Joint Research
program "Mesoscopic Electronics" and by Grant-in-Aid for Scientific Research
on Priority Area, "Electron Wave Interference Effects in Mesoscopic Structure"
(04221101) and by Monbusho International Scientific Research Program:Joint
Research "Theoretical Studies on Strongly Correlated Electron Systems"
(05044037) from the Ministry of Education, Science and Culture of Japan.

\begin{figure}

\caption{
Configuration of a planar domain wall.
  \label{DW} }

\caption{
Strength of ohmic dissipation $\eta$ given by Eq. (\protect\ref{eta})
as a function of the width of the wall $\lambda/a$, $a$ being a lattice
constant, for three choices of $\delta\equiv (\kfu-\kfd)/(\kfu+\kfd)=0.05,
0.1$ and $0.2$ with $(\kfu+\kfd)a =6.0$.
  \label{FIGeta} }

\caption{
The tunneling rate $\Gamma$ for the insulating ($\eta=0$) (dashed line)
and the metallic ($\eta=0.1$) (solid line) magnet as a function
of $\epsilon$, $H_c$ being the coercive field. Number of spin is
$N=10^4$ and $10^8$.
  \label{FIGgamma} }

\end{figure}


\begin{thebibliography}{99}
\bibitem{aniso} The results are applicable also to the case of uniaxial
anisotropy if $K_\perp$ is replaced by the demagnetization energy
$\mu_0(\hbar\gamma)^2/a^6$.
\bibitem{RPA} The RPA to the interaction, $U$, is taken
because we have not included in the magnetization
 the Coulomb interaction with finite momentum transfer
$\Huf\equiv -({U}/{2})$$\sum_{\qv\neq0 }$$
\sum_{\kv\kv'}$$
(a^\dagger_{\kv+\qv} \sigma_z a_\kv )(a^\dagger_{\kv'-\qv} \sigma_z a_{\kv'} )
$.
\end{thebibliography}
\end{document}